# Professional Ethics by Design: Co-creating Codes of Conduct for Computational Practice

Samuel Danzon-Chambaud*

frog, part of Capgemini Invent, samuel.danzon-chambaud@frog.co

Marguerite Foissac

frog, part of Capgemini Invent, marguerite.foissac@frog.co

This paper deals with the importance of developing codes of conduct for practitioners—be it journalists, doctors, attorneys, or other professions—that are encountering ethical issues when using computation, but do not have access to any framework of reference as to how to address those. At the same time, legal and technological developments are calling for establishing such guidelines, as shown in the European Union's and the United States' efforts in regulating a wide array of artificial intelligence systems, and in the resurgence of rule-based models through "neurosymbolic" AI, a hybrid format that combines them with neural methods. Against this backdrop, we argue for taking a design-inspired approach when encoding professional ethics into a computational form, so as to co-create codes of conduct for computational practice across a wide range of fields.

## 1 INTRODUCTION

In 2019, a team of journalists and developers at the BBC worked on setting up automated news—that is, automated text generation for journalistic purposes [7] [17]—to cover the results of the general election in the United Kingdom. In doing so, they were faced with a somewhat unconventional challenge: having to delineate the exact journalistic rules that would go into the algorithm behind automated news [12]. For instance, this could involve reflections on determining a threshold to qualify the magnitude of a win. By how many votes should we tell the algorithm that it is "a large victory" as opposed to a "narrow win"? Likewise, below which percentage can it be considered a crushing defeat?

In journalism like in any other practice governed by professional ethics (for instance healthcare or law practice), work-related interrogations that are encountered on a day-to-day basis are usually addressed *via* professional codes of conduct, which give recommendations as to a preferred way of acting in a given set of circumstances (e.g., the BBC's Editorial Guidelines, the American Medical Association's Code of Medical Ethics, bar associations' codes of conduct); however, these recommendations generally concern "real-life" situations and are not applicable the realm of algorithms and computation. Albeit not directly related to professional codes of conduct, a close enough example where computation and professional ethics intermingle is the Handbook of Sustainable Design of Digital Services (GR491, which frog's parent company, Capgemini, was involved in [25]) launched by the French Institute for Sustainable IT (Institut du numérique responsable or INR). A set of 516 criteria that help digital professionals like project managers, front-end developers or interaction designers with reducing the environmental and social repercussions of the digital services they set up [20], GR491 has been collectively conceived by a working group of volunteers that included experts in UX design and sustainability.

---

\* First and corresponding author

Given the overwhelming role taken by datafication and information processing today—which brings about democratic concerns like filter bubbles [32], surveillance [41] and race and gender considerations [15]—we believe it is critical to reflect on computational aspects that are not yet a part of professional ethics. In this position paper, we first detail the legal and technological context that calls for developing professional codes of conduct for computational practice, then suggest a design-inspired approach that would be best suited to developing those. Ultimately, our goal is to bring forth a standardized set of procedures that can be emulated across disciplines, and where practitioners remain at the center of it.

## 2 LEGAL AND TECHNICAL CONTEXT

Following the adoption of the General Data Protection Regulation (GDPR) that set limits to the business of datafication, the European Union has undertaken to be regulating AI so that it is safe for users and complies with existing laws [10]. The General Approach adopted at the end of 2022 is quite extensive in length: it is inclusive of rule-based and machine learning systems and takes a risk-based approach to regulating artificial intelligence. The main criticism, though, generally involves the idea that it would slow down innovation, as reported by a group of AI associations [26]. By contrast, the United States' Blueprint for an AI Bill of Rights, unveiled in October 2022, features a few guiding principles followed by a much more detailed Technical Companion, which provides guidance as to how to implement them. The Blueprint exhibits a more all-encompassing definition of AI than the EU's, yet shares similar concerns on preserving fundamental rights like safety for users [38]. That said, the US approach has been criticized for "lacking teeth" as it misses out on enforcement aspects [18] [21] [11].

Most interesting to us here are two dispositions contained in EU's General Approach and in the US' Blueprint: first, Article 69 of the General Approach specifically refers to the drawing up of codes of conduct that relate, among others, to "stakeholders participation in the design and development of the AI systems"; second, in its recommendations for "Safe and Effective Systems", the Technical Companion to the Blueprint stresses the importance of having "early-stage consultation" with impacted communities, but also with relevant stakeholders like "subject matter, sector-specific, and context-specific experts". Taken together, this testifies of lawmakers' inclination for an "Ethics by Design" approach to regulating AI—that is, when ethical considerations are thought of well ahead of writing the first line of code [16]. Our suggestion for establishing professional codes of conduct for computational practice therefore seems timely and relevant, as it would enable professional ethics to be embedded in the design of AI systems.

Besides complying with this regulatory background, coming up with professional codes of conduct for computational practice could also constitute an asset in the development of "neurosymbolic" AI systems, which may call for increased attention to encoding expert knowledge into algorithmic rules. Neurosymbolic systems generally refers to a blend of the two conflicting positions that have dominated much of the history of AI: on the one hand, the symbolic approach advocated by McCarthy, Minsky, Simon and Newell—which aimed at translating a wide vision of human understanding into computer code—and on the other hand, connectionist methods brought forth by Rosenblatt and, later, Lecun, Hinton and Bengio—which are rather focused on churning through large amounts of data using one or several layers of artificial "neurons" that mimic the functioning of the brain, relying for that on an activation function. Despite being the prevailing perspective up until the 1990s and leaving people in awe when IBM's Deep Blue beat chess champion Garry Kasparov, symbolic AI has since waned off to give way to connectionist approaches [8]: those have, indeed, made impressive breakthroughs in recent years, especially in the realm of computer vision as shown in Krizhevsky, Sutskever and Hinton's groundbreaking neural architecture [24]. However, connectionism has also been criticized for, among others, being too opaque and, by extent, too unpredictable, thus reinforcing the idea that neurosymbolic systems will be at the forefront of AI research in the coming years [28] [29] [1].

According to Kautz' taxonomy [22], as of late significant AI breakthroughs—including some large language models—have incorporated a neurosymbolic dimension: for example, this is the case of Google's subsidiary DeepMind that beat world champion Lee Sedol at the very complex game of Go [34], of a software that learns by "looking at images and reading paired questions and answers" similarly to the way a human does [27, p. 1] and, more recently, of a program developed by Meta that, as pointed out by Marcus and Davis [30], makes use of neurosymbolic elements to play the highly reflective game Diplomacy at human-level performance [31]. Even though much work remains to be done on the best way forward to integrate connectionist and symbolic approaches [2] [22], this is nonetheless clear evidence of the resurgence of rule-based methods under this hybrid format: as such, professional codes of conduct for computational practice could be critical to encoding professional ethics into a new generation of neurosymbolic AI systems. As we will show next, a human-centric take on design that is grounded in social responsibility appears to be the most relevant means to creating these guidelines.

## 3 THE NEED FOR A DESIGN-INSPIRED APPROACH

As a discipline that has become increasingly oriented toward co-creation [6], design can be seen as the most appropriate framework to work with to come up with professional codes of conduct for computational practice. At its core, design takes a holistic look at the relationship between designed artifacts, people that are exposed to these artifacts and associated social, cultural and business contexts: as such, it primarily pays attention to user experience as its original focus is on valuing users first and foremost. That said, there is growing consensus now that design can also lead the way toward social good as it provides a unique set of skills, tools and methods that can be used in that regard [5]. According to Tromp, Hekkert and Verbeek [35], the idea is that design transcends problems that relate to usability only (i.e., user-centered design) to deal with questions that address social change as well (i.e., human-centered design). Thus, *social design* or *design for social innovation*—as this vision is called [36] [14]—is much relevant in a domain like professional ethics where the goal is to foster common good. Going into the specifics, we could follow one of the many design methodologies that are available to date [23], and which generally start with a phase of immersion and understanding (e.g., observations, interviews, surveys), followed by a key phase of ideation that can most notably be done through co-conception, before switching to a prototyping phase and, finally, a stage of testing and evaluation.

Relevant to co-conception here are some of the attributes provided by *game design*, another design stream with connections to two of the most prominent views on the act of designing: one carried by Simon (also mentioned above in the context of symbolic AI), whose focus is on rational problem-solving through decomposition [19] [40]; the other advocated by Schön, according to whom reflection and action feed into each other in a sort of iterative loop [37]. Close to Simon's ideas is Bjork, Lundgren and Holopainen's [4] inventory of over 200 game design blocks or "patterns" (e.g., "paper rock scissors"), whereas Bateman and Boons' mention of Japanese game designers—which, according to them [3, p. xii], "display a kind of holistic thinking that defies decomposition into method" — sits closer to Schön's argument, but is harder to encapsulate into a workable model [9]. During the co-conception phase, we could then draw on a Simon-inspired approach to game design so as to encourage practitioners to come up with their own ideas on how to translate professional ethics into a computational format, which also presents the advantage of sharing the same epistemological roots.

## 4 CONCLUSION

In this paper, we detailed the legal and technological context that calls for developing professional codes of conduct for computational practice: first, the European Union's proposed AI Act and the United States' Blueprint for an AI Bill of Rights are both giving weight to an Ethics by Design approach where stakeholders are involved in the conception and development of AI systems, which shows the importance of encoding professional ethics into a computational form from the very beginning; second, the resurgence of rule-based models through neurosymbolic AI makes for having such guidelines already established, in order to integrate those into the development of future systems. What is more, we have also made the case for taking a social design perspective when conceiving these professional codes of conduct, which can be reinforced by a Simonian approach to game design in the co-conception phase.

The strength of using such a design-inspired approach rests in that it is not limited to any one domain, but—on the contrary—that it can be extended to many fields of application, like professional ethics within the journalistic, medical or legal community. Additionally, it brings to light another core issue, which is translating the specifics of any real-life situation into a form of abstraction that is suitable to a computational format, for which a proper design methodology is yet to be developed. If anything, New Zealand's efforts in rendering the law into the form of computer code shows the importance of undertaking this [13]: the program's ambition is to interact with businesses' and individuals' software so that it helps them better understand regulations and thus comply with it. This is very much line with Wing's observation on "computational thinking" [39], a way of solving problems, designing systems, and understanding human behavior that has abstraction and decomposition at its heart, but which also carries the complex and sometimes opaque task of translating human semantics into computer syntax [33].

## ACKNOWLEDGMENTS

We are grateful to frogLab's director, Clément Bataille, and to our colleagues Rose Dumesny, Benjamin Martin and Yasmine Saleh for their insights and encouragement.